\documentclass[twocolumn,showpacs,preprintnumbers,prl,fleqn,superscriptaddress]{revtex4}
\usepackage{graphicx}
\usepackage{dcolumn}
\usepackage{bm}
\usepackage{natbib}

\begin{document}

\title{Magneto-Raman scattering of graphene on graphite: Electronic and phonon excitations}

\author{C. Faugeras} \affiliation{LNCMI, UPR 3228, CNRS-UJF-UPS-INSA, 38042 Grenoble, France}
\author{M. Amado}
\affiliation{LNCMI, UPR 3228, CNRS-UJF-UPS-INSA, 38042 Grenoble,
France} \affiliation{QNS-GISC, Departamento de F\'{i}sica de
Materiales, Universidad Complutense, E-28040 Madrid, Spain}
\author{P. Kossacki}
\affiliation{LNCMI, UPR 3228, CNRS-UJF-UPS-INSA, 38042 Grenoble,
France} \affiliation{Institute of Experimental Physics, Faculty of
Physics, University of Warsaw, Poland.}
\author{M. Orlita}
\affiliation{LNCMI, UPR 3228, CNRS-UJF-UPS-INSA, 38042 Grenoble,
France} \affiliation{Institute of Physics, Faculty of Mathematics
and Physics, Charles University, Ke Karlovu 5, 121 16 Praha 2,
Czech Republic}
\author{M. K\"{u}hne}
\affiliation{LNCMI, UPR 3228, CNRS-UJF-UPS-INSA, 38042 Grenoble,
France}
\author{A.A.L. Nicolet}
\affiliation{LNCMI, UPR 3228, CNRS-UJF-UPS-INSA, 38042 Grenoble,
France}
\author{Yu. I. Latyshev}
\affiliation{Kotelnikov Institute of Radio-Engineering and
Electronics RAS, Mokhovaya 11-7, 125009 Moscow, Russia}
\author{M. Potemski}
\affiliation{LNCMI, UPR 3228, CNRS-UJF-UPS-INSA, 38042 Grenoble,
France}
\date{\today }

\begin{abstract}

Magneto-Raman scattering experiments from the surface of graphite
reveal novel features associated to purely electronic excitations
which are observed in addition to phonon-mediated resonances.
Graphene-like and graphite domains are identified through
experiments  with $\sim 1\mu m$ spatial resolution performed in
magnetic fields up to 32T. Polarization resolved measurements
emphasize the characteristic selection rules for electronic
transitions in graphene.  Graphene on graphite displays the
unexpected hybridization between optical phonon and symmetric
across the Dirac point inter Landau level transitions. The results
open new experimental possibilities - to use light scattering
methods in studies of graphene under quantum Hall effect
conditions.
\end{abstract}
 \maketitle

Effects of interactions, between electrons and/or electrons and
phonons, turn out to be an emergent focus of the research on
graphene~\cite{Geim2009}. These effects are particularly apparent
when the magnetic field $B$ is applied across the graphene plane
and the continuous energy dispersion spectrum $E= \pm v_{F}\mid
\overrightarrow{p} \mid$ of this Dirac, two-dimensional system
transforms into discrete Landau levels $L_{\pm n}$ with
characteristic energies $E_{\pm n} = \pm v_{F} \sqrt{2 e \hbar B
n}$ (where \overrightarrow{p} is carrier momentum, $n=0,1, .., $
and $v_{F}$ stands for the Fermi velocity). Magneto-Raman
scattering is one, possible spectroscopic tool to study the
(selected) inter Landau level excitations, which is potentially
sensitive to the effects of
interactions~\cite{Pinczuk1992,Vankov2010}.

So far, a search for Raman scattering signal from (purely)
electronic excitations between LLs of graphene
systems~\cite{Kashuba2009,Roldan2010,Mucha-Kruczynski2010} has
been a veritable experimental challenge~\cite{Garcia09}. The
dominant electronic resonances are predicted to be associated with
the symmetric inter LL transitions ($L_{-n}\rightarrow L_{n}$)
across the $0^{th}$ Landau level in the case of
graphene~\cite{Kashuba2009,Roldan2010} and of bilayer
graphene~\cite{Mucha-Kruczynski2010}. Notably, however, the
resonant hybridization of the $E_{2g}$ optical phonon
mode~\cite{Ando07}, with asymmetric $L_{-n, (-n-1)} \rightarrow
L_{n+1,(n)}$ excitations~\cite{Goerbig07}, has been
observed~\cite{Faugeras09,Yan2010} and displayed a characteristic
nature of electron-phonon interaction in graphene. If exfoliated
graphene samples are often limited in quality or very fragile, the
extremely pure graphene specimens which can be found on the
surface of bulk graphite~\cite{Li2009,Neugebauer2009} might be
more suitable for the refined spectroscopic studies~\cite{Yan2010}
and investigations of both electronic and phonon-mediated response
in magneto-Raman scattering experiments.

In this Letter, we report on macro and micro magneto-Raman
scattering experiments performed on natural graphite. Our results
reveal a series of characteristic transitions which we assign as
due to the electronic response and which are observed in addition
to previously studied, phonon-mediated excitations. A mixture of
graphene and graphite response is seen in macro experiments. By
scanning the sample surface with micro-Raman scattering at high
fields, we localize the graphene flakes on graphite
substrate~\cite{Yan2010} via mapping the $E_{2g}$ magneto-phonon
resonance. Focusing on such a location we experimentally identify
the dominant features in the electronic Raman scattering response
of graphene, which, in agreement with theoretical predictions,
arise from symmetric, across the Dirac point, inter Landau level
excitations. Strikingly and beyond the existing theoretical
models, these symmetric excitations also couple to optical phonon
$E_{2g}$ mode. Electron-electron interactions and/or symmetry
breaking effects due to interaction with the substrate are
speculated to be among possible causes of this effect.

Raman scattering spectra were measured using the Ti:Saphire laser
setup, tuned at accurately controlled wavelength in the range
$\sim\lambda$=720~nm, in order to minimize the superfluous Raman
signal of optical fibers. The sample was immersed in a helium gas
kept at T$=4.2$~K and placed in a resistive magnet delivering
fields up to $32$~T. The non-polarized Raman scattering spectra
were measured in nearly back-scattering Faraday geometry. The
collected light was dispersed with a single grating spectrometer
(spectral resolution $\Delta\lambda$=0.3~nm) equipped with
nitrogen cooled CCD detector and band pass filters were used to
reject the stray light. For macro-Raman scattering experiments, an
experimental set-up similar to the one described in
Ref.~\cite{Faugeras09} was used with a laser spot of 600~$\mu$m
-diameter and an optical power $\sim 100$~mW. For micro-Raman
measurements, a mono-mode fiber with a core of 5~$\mu$m diameter
was used for optical excitation and a 200~$\mu$m core optical
fiber was used for collection. Aspherical lenses were used for
excitation and collection. The laser spot on the sample was $\sim$
1~$\mu$m and an optical power of 5 mW was used for excitation. The
sample has been mounted on a X-Y-Z piezzo-stages allowing to move
the sample with respect to the laser spot with a spatial
resolution better than 1~$\mu$m.

\begin{figure}
\scalebox{0.4}{\includegraphics*{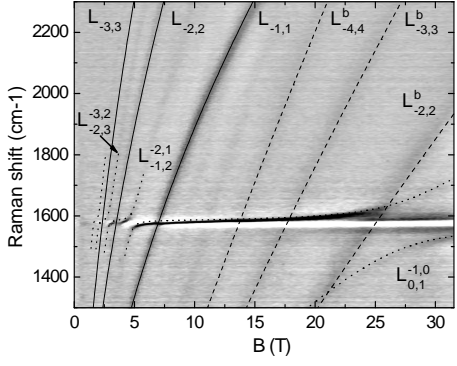}}
\caption{\label{Fig1} False color map of the B=0T spectrum
subtracted macro-Raman spectra of natural graphite as a function
of the magnetic field. Dotted lines are the graphene-optical
phonon coupled modes discussed in Ref.~\cite{Goerbig07}, solid
lines are the calculated symmetric inter Landau level transitions
L$_{-n,n}$ in graphene, dashed lines are the symmetric transitions
L$^{b}_{-n,n}$ in the effective bilayer. All these lines are
calculated with a Fermi velocity $v_{F}=1.02\cdot10^{6}
$m.s$^{-1}$ and $2\gamma_{1}=750$ meV.}
\end{figure}

The results of our macro-Raman scattering measurements are
summarized in Fig.~\ref{Fig1} which shows a false color plot of
spectra with subtracted B=0 response, measured in the spectral
range of the vicinity of the E$_{2g}$ optical phonon energy, as a
function of the magnetic field. Different series of magnetic field
dependent features can be distinguished, with different
intensities and different behaviors when tuned in resonance with
the E$_{2g}$ optical phonon energy.

Two, $L_{-n,n}$ and $L_{-n,n+1}^{-n-1,n}$ series of features scale
with $\sqrt{B}$ and can be associated with transitions which
involve Landau levels $L_{n}$ of massless Dirac fermions with
Fermi velocity $v_{F}=1.02\cdot10^{6}$m.s$^{-1}$. Better
pronounced $L_{-n,n}$ series is due to symmetric,
$L_{-n}\rightarrow L_{n}$ transitions whose energies are
reproduced with the calculated solid lines in Fig.~\ref{Fig1}.
These transitions are expected to dominate the electronic
magneto-Raman scattering response of graphene~\cite{Kashuba2009}.
The overall weaker, $L_{-n,n+1}^{-n-1,n}$ series is due to
asymmetric $L_{n,(-n-1)}\rightarrow L_{n+1,(n)}$ LL transitions
(optical-like excitations) which, on the other hand, give rise to
the pronounced magneto-phonon resonance. Dotted lines in
Fig.~\ref{Fig1} represent the coupled, optical phonon-electronic
excitations as defined in Ref.~\cite{Goerbig07} for neutral
graphene and calculated with an electron-phonon coupling constant
of 3000$\pm$50 cm$^{-1}$~\cite{Yan07,Faugeras09}.

The third, $L^{b}_{-n,n}$ series of features which scale more
linearly with the magnetic field is here attributed to Raman
scattering response arising from massive electrons at the K-point
of graphite~\cite{Slonczewski58,McClure56}. The in plane
dispersion of these electronic states can be well approximated by
the one of the effective bilayer graphene with the interlayer
coupling $\gamma_{1}$ enhanced by a factor of two with respect to
a true bilayer
graphene~\cite{Koshino2008,Orlita08a,Orlita2009,Chuang2009}. The
$L^{b}_{-n,n}$ features can be therefore identified with the
symmetric $L^{b}_{-n} \rightarrow L^{b}_{n}$ transitions between
LL levels of such an effective bilayer graphene. Energies $E^{b}
_n$ of these levels are given by:
\begin{eqnarray}
E^{b}_n & = &
sgn(n)\frac{1}{\sqrt{2}}\Big{[}(2\gamma_{1})^{2}+(2n+1)E_{1}^{2}\nonumber\\
     & -&
     \sqrt{(2\gamma_1)^2+2(2n+1)E_{1}^{2}(2\gamma_1)^2+E_{1}^4}\Big{]^{1/2}}
\nonumber
\end{eqnarray}

where $E_{1}=v_{F}\sqrt{2 e \hbar B}$ and energies of $L^{b}_{-n}
\rightarrow L^{b}_{n}$ transitions are $2E^{b}_{n}$. Those latter
values have been calculated assuming that $\gamma_1=375$ meV (and
$v_{F}=1.02\cdot10^{6}$m.s$^{-1}$) and are plotted in
Fig.~\ref{Fig1} with dashed lines which coincide well with the
measured $L^{b} _{-n,n}$ transitions. Although we are not aware of
any theoretical works on magneto-Raman scattering response of bulk
graphite, the predictions regarding the bilayer graphene are in
line of our identification of the $L^{b}_{-n} \rightarrow
L^{b}_{n}$ excitations. Notably those transitions are very likely
the same as previously reported~\cite{Garcia09} but interpreted
differently.

\begin{figure}
\scalebox{0.4}{\includegraphics*{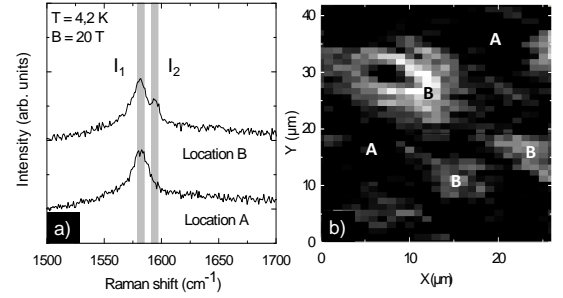}} \caption{\label{Fig2}
a) Unpolarized micro-Raman scattering spectra at B=20T for two
different representative locations on the surface of natural
graphite. The shaded area represent the energy interval from which
the average amplitude $I_1$ and $I_2$ are extracted. b) False
color map of the intensity ratio I$_1$/I$_2$ showing different
regions on the surface of natural graphite. At B = 20 T, the
$E_{2g}$ phonon appears as a single component feature in locations
\emph{A}, as a two components feature in locations \emph{B}}
\end{figure}

More striking is our observation of the series of the $L$-type,
graphene-like transitions. One could speculate that these
transitions arise due to electronic states at the H-point of
graphite, which to some extend also reflect the behavior of
massless Dirac fermions~\cite{Orlita2009}. The origin of these
transitions is however different as we elucidate this with micro
magneto-Raman scattering experiments. In Fig.~\ref{Fig2}a we
present two micro-Raman scattering spectra, both measured at B=20T
but at two different locations on the surface of natural graphite.
When the laser spot is at location \emph{B}, the E$_{2g}$ optical
phonon appears clearly as a two-components feature indicating an
efficient electron-phonon coupling, similar to the one discussed
in Ref.~\cite{Faugeras09}. In contrast, at the location \emph{A},
the E$_{2g}$ optical phonon appears as a single component feature
pointing towards much weaker effect of electron-phonon coupling.
Let us define I$_1$ as the amplitude of the scattered light at the
E$_{2g}$ optical phonon energy and I$_2$ as the amplitude of the
scattered light at the energy where the second component appears
(see Fig.~\ref{Fig2}a). In Fig.~\ref{Fig2}b we present a
false-color map of the ratio R=I$_1$/I$_2$ as a function of the
position of the excitation spot on the surface of the sample. In
regions of high R (dark regions in Fig.~\ref{Fig2}b, the E$_{2g}$
optical phonon appears as a single component feature (\emph{A}
spectrum in Fig.~\ref{Fig2}a) while low R indicates regions where
the E$_{2g}$ optical phonon appears as a double component feature
(light gray regions labelled \emph{B} in Fig.~\ref{Fig2}b). The
surface of natural graphite is therefore not homogeneous.
\emph{A}-regions display predominantly the excitations
characteristic of massif carriers at the K point of the band
structure of bulk graphite. The magneto-Raman spectra from such
locations will be discussed in more details elsewhere.

\begin{figure}
\scalebox{0.35}{\includegraphics*{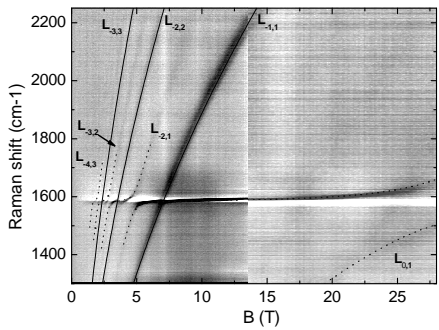}} \caption{\label{Fig3}
False color map of the B=0T spectrum subtracted micro-Raman
spectra measured with the laser spot on location \emph{B} as a
function of the magnetic field. Solid lines correspond to
symmetric transitions L$_{-n,n}$ in graphene while dotted lines
correspond to the hybrid modes involving E$_{2g}$ phonons and
optical-like transitions in graphene.}
\end{figure}

Here we focus on \emph{B}-locations, of few square micrometers,
which we identify as graphene inclusions (decoupled graphene
flakes) on the surface of the bulk
graphite~\cite{Li2009,Neugebauer2009}. Indeed, as shown in
Fig.~\ref{Fig3}, the Raman scattering response measured with
micron resolution from the \emph{B}-location is practically free
of $L^{b}$ graphite features. The remaining L-type transitions are
characteristic of graphene and  are not observed when placing the
laser spot on \emph{A}-locations. Solid and dotted lines in
Fig.~\ref{Fig3} are the corresponding traces redrawn from
Fig.~\ref{Fig1} and they reflect well the overall behavior of the
observed transitions, which can be now analyzed in more details.

\begin{figure}
\scalebox{0.35}{\includegraphics*{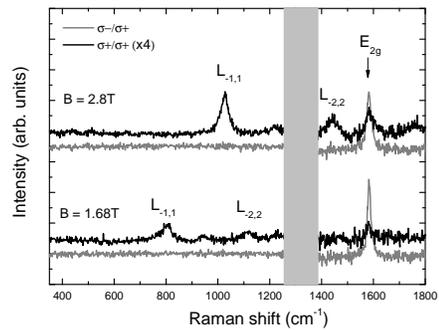}} \caption{\label{Fig4}
Micro-Raman scattering spectra measured at a \emph{B}-location at
B=1.68T and 2.8T at T=4.2K in the $\sigma+/\sigma-$ (gray curves)
and in the $\sigma+/\sigma+$ (black curves x4) polarization
configurations (excitation/detection) showing two purely
electronic Raman scattering features L$_{-1,1}$ and L$_{-2,2}$.
The shaded region is due to a magnetic field dependent background
feature affecting the spectra.}
\end{figure}

To further elucidate the nature of the electronic L$_{-n,n}$
transitions, we have performed polarization resolved experiments
at low magnetic fields. As shown in Fig.~\ref{Fig4}, they can be
observed far away from the E$_{2g}$ phonon resonance. Moreover,
the L$_{-n,n}$ symmetric transitions imply no change of angular
momentum and, in accordance to theoretical
predictions~\cite{Kashuba2009}, are seen when the helicity of the
circularly polarized incoming light and of the outgoing scattered
signal are the same. In contrast, the optical phonon with the
E$_{2g}$ symmetry~\cite{Goerbig07,Kashuba2009} dominates the
spectra in crossed polarization configuration. The relative
intensity of these electronic features with respect to the one of
the phonon feature strongly depends on the location on the sample.

\begin{figure}
\scalebox{0.33}{\includegraphics*{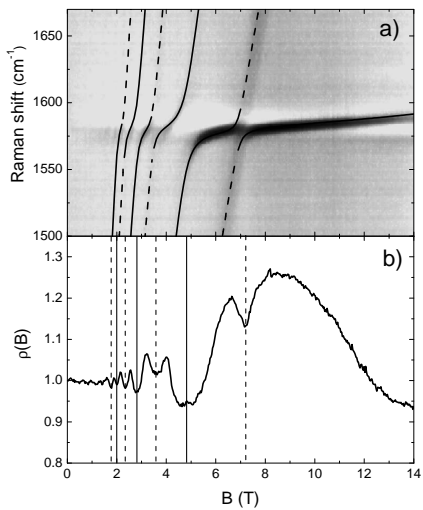}} \caption{\label{Fig5}
a) False color map of the B=0T spectrum subtracted Raman
scattering spectra in the E$_{2g}$ phonon range of energy. The
solid and dashed lines represent optical and symmetric electronic
excitations coupled to optical phonons. b) $\rho$(B), amplitude of
the Raman scattering response at the E$_{2g}$ phonon energy
(1582.2 cm$^{-1}$) normalized to the zero field amplitude, as a
function of magnetic field. Solid lines represent the resonant
fields of optical-like transitions with the E$_{2g}$ phonon while
dotted lines are the resonant fields of symmetric transitions with
the E$_{2g}$ phonon.}
\end{figure}

Finally, let us turn our attention on the close vicinity of the
E$_{2g}$ phonon signal and focus carefully on its behavior at low
magnetic fields, below~8T. Fig.~\ref{Fig5}a shows a false color
map of the unpolarized Raman spectra with subtracted B=0 spectrum
and Fig.~\ref{Fig5}b illustrates the field dependence of
$\rho(B)$, the amplitude of the Raman scattering signal at the
energy which corresponds to the center of the $E_{2g}$ peak
measured at zero magnetic field (1582.2 cm$^{-1}$) and scaled with
respect to its zero field value. As expected, this trace shows
minima, marked by solid lines in Fig.~\ref{Fig5}a, which appear at
magnetic fields at which the energies of $L_{-n-1,(-n)}\rightarrow
L_{n,(n+1)}$ transitions coincide with the $E_{2g}$-phonon energy.
These minima are the result of regular magneto-phonon
effect~\cite{Ando07,Goerbig07,Faugeras09,Yan2010} which \textrm{de
facto} implies an avoided crossing between those asymmetric LL
transitions and the optical phonon mode. Surprisingly, the
$\rho(B)$ dependence shows another set of well pronounced minima,
marked by dashed lines in Fig.~\ref{Fig5}b (The behavior of
$\rho(B)$ above 8T is reflecting the shift of the phonon line
towards higher energies). They clearly appear when the symmetric
$L_{-n}\rightarrow L_{n}$ transitions are tuned in resonance with
the $E_{2g}$-phonon. We therefore deduce and illustrate it more
clearly in Fig.~\ref{Fig5}b that both asymmetric and symmetric
inter Landau level transitions show the effect of avoided crossing
with the $E_{2g}$ phonon mode. To be more quantitative, we have
simulated the coupling of both asymmetric and symmetric LL
transitions to the phonon mode (see Fig.~\ref{Fig5}a). In case of
asymmetric transitions, we have repeated the calculations
according to the prescription of Ref.~\cite{Goerbig07}, in which
the coupling parameters for each resonance grow with magnetic
field and are determined by the strength (3000$\pm$50 cm$^{-1}$)
of electron-phonon interaction. In case of symmetric transitions
we have phenomenologically introduced the coupling parameters,
separately for each resonance, and found $\delta=4.7, 11.1,$ and
$15.7$ cm$^{-1}$ for the resonances of the $E_{2g}$-phonon with
correspondingly $L_{-1,1}$, $L_{-2,2}$, and $L_{-3,3}$ transitions
which appear at $B=7.19, 3.61$, and $2.31$ T, respectively. The
decreasing with magnetic field strength makes us speculate that it
originates from the mixing of wave functions of neighboring Landau
levels. Such mixing may result from Coulomb, electron-electron
interactions~\cite{Roldan2011}. It is however surprising that the
couplings involving symmetric lines observed at low magnetic
fields, are as strong as the standard magneto-phonon effect.
Notably, graphene on graphite may consist of rotationally twisted
sheet with respect to Bernal-stacked underneath layers. The role
of substrate, and possible other symmetry breaking effects cannot
be \textit{a priori} excluded. We believe our observation will
trigger pertinent theoretical studies which are beyond the scope
of our experimental paper.

Concluding, we have presented low temperature macro and
micro-Raman scattering experiments in magnetic fields performed on
natural graphite, which show the existence of graphene flakes
decoupled from bulk graphite and which reveal the rich electronic
excitation spectrum of graphene. We have determined the different
polarization selection rules for these excitations. The
magneto-phonon effect of undoped graphene is revealed as well as a
coupling of electronic transitions symmetric across the Dirac
point and the optical phonon. This coupling is observed because of
the mixing of Landau levels wave functions. The origin of this
mixing remains unknown what calls further theoretical works on
electron-phonon coupling in graphene. Our observation of
electronic response in magneto-Raman scattering from graphene
opens new possibilities to studying the properties of this new
quantum Hall effect system.

\begin{acknowledgments}
We would like to acknowledge fruitful discussions with D.M. Basko
and V.I. Falko. Part of this work has been supported by
ANR-08-JCJC-0034-01, GACR P204/10/1020, GRA/10/E006 (EPIGRAT),
RTRA "DISPOGRAPH" projects and by EuroMagNET II under the EU
contract number 228043. P.K. is financially supported by the EU
under FP7, contract no. 221515 ''MOCNA''. Yu. L. is supported by
the Russian state contract No. 16.740.11.0146.
\end{acknowledgments}

\end{document}